\documentclass{pasa}%

\title[Bulge 3D Structure]{The 3D structure of the Galactic bulge}
\author[Zoccali \& Valenti]{Manuela Zoccali$^{1,2}$ \and Elena Valenti$^3$ \\
\affil{$^1$Instituto de Astrof\'isica, Pontificia Universidad  Cat\'olica de Chile, 
Av. Vicu\~na Mackenna 4860, 782-0436 Macul, Santiago, Chile}
\affil{$^2$Millennium Institute of Astrophysics, Av. Vicu\~na Mackenna 4860, 
782-0436 Macul, Santiago, Chile}
\affil{$^3$European Southern Observatory, Karl-Schwarzschild Strasse 2, D-85748 
Garching, Germany}}
\jid{PASA}
\doi{10.1017/pas.\the\year.xxx}
\jyear{\the\year}

\usepackage[authoryear]{natbib}
\bibpunct{(}{)}{;}{a}{}{,}
\begin{document}

\begin{abstract}
We review  the observational  evidences concerning the three-dimensional structure  of the
Galactic bulge.  Although  the inner few kpc  of our Galaxy are normally  referred to as
{\it the  bulge}, all the  observations demonstrate that this  region is dominated  by a
bar, i.e., the bulge  is a bar. The bar has a boxy/peanut  (X\--shaped) structure in its
outer regions, while it seems to become less and less elongated in its innermost region.
A thinner and longer structure departing from the main bar has also been found, although
the observational  evidences that support  the scenario  of two separate  structures has
been  recently  challenged.   Metal  poor  stars ([Fe/H]$\lesssim  -0.5$  dex)  trace  a
different structure, and also have different kinematics.
\end{abstract}


\maketitle

\section{INTRODUCTION}
\label{sec:intro}

In the last  few years Galactic archaeology  has risen as a branch  of astrophysics that
seeks to unveil the origin of a stellar  system by studying fossil records in its oldest
stars.  The oldest stars of the Milky Way have been found in the halo. It is in the halo
where we can find traces of the  Galaxy's primordial components, including streams of accreted
satellites.  The halo,  however, makes up only  about ~1\% of the total  stellar mass of
the Milky Way  \citep[$<10^9$ M$_\odot$;][]{robin+03,bell+08}. If we  seek to understand
how the  bulk of  the Milky  Way formed,  a valuable  alternative is  to study  the best
compromise between old and massive, that is the Galactic bulge.

As reviewed elsewhere in  this volume (e.g., paper by Gerhard) the  mass of the Galactic
bulge   is  rather   poorly   constrained.   Most   determinations   cluster  close   to
$1.5\times10^{10}$  M$_\odot$,  although   a  few  authors  find  values   as  large  as
$2\times10^{10}$  M$_\odot$ \citep{valenti+15}  or as  small as  $6\times10^9$ M$_\odot$
\citep{robin+12}. Even  with this rather  large scatter, the mass  of the bulge  must be
close to 1/5 of the total stellar mass of the Milky Way, and about ten times larger than
the mass of the halo.  The bulge age distribution is also currently debated, but all the
studies   agree   that   the   bulk   of    bulge   stars   is   $\sim   10$   Gyr   old
\citep{ortolani+95,kuijken+02,zoccali+03,sahu+06,clarkson+11,valenti+13},           with
discrepancies regarding only  the presence, and the number fraction,  of an intermediate
age tail of the distribution \citep[see e.g.][]{bensby+11,bensby+13}.

In addition to being  old and massive, the Galactic bulge is the  only galaxy bulge that
can be  resolved down  to its faintest  stars: a  unique case that  can be  studied with
exquisite  details. 

\cite{kormendy+04}  classified  galaxy  bulges  as   {\it  classical}  bulges  and  {\it
  pseudo-}bulges.  Classical bulges would be spheroids formed by gravitational collapse,
or hierarchical  merging of  smaller galaxies. They  would be formed  in the  very early
epoch of galaxy formation,  and thus are usually older than  the disk. Pseudo-bulges, on
the contrary, would  be smaller, disk-like structures found in  the innermost regions of
spirals, and originated because the presence of  bars favors the accumulations of gas in
the very  inner part of  the disks, forming  either an inner disk  or a ring.   Once the
disk/ring is sufficiently massive it starts forming  stars that can be detected as young
population in the central part of the  disks.  This classification has been used widely,
often with the over simplification that  spheroids would be classical bulges, while bars
would be  pseudo-bulges, because they  are originated from disk  instabilities.  Through
this paper  we will abandon this  classification, mostly because recent  observations of
high redshift  galaxies suggest  that the  origin of  bulges can  be much  more complex,
likely  related to  the  merging of  dense,  star  forming clumps  present  in the  disk
\citep{immeli+04, carollo+07, elmegreen+08,  bornaud+09}. Hence in what  follows we will
refer to the ``bulge'' as the Milky Way region within a radius of $\sim$2-3 kpc from the
Galactic center, without any implication on its nature or origin.

This paper  reviews the progress  of our knowledge  of the bulge  three-dimensional (3D)
structure, focusing  on the  observational results. We  occasionally present  an updated
version of  the relevant figures,  made using the state  of the art  observational data.
Section~\ref{sec:mainbar}  presents evidences  for the  presence of  a main  bar in  the
central region of  the Milky Way. Section~\ref{sec:xshape} explains how  we became aware
of     a     boxy/peanut    (i.e.,     X\--shaped)     structure     in    the     outer
bulge.  In Section~\ref{sec:spheroid}  we discuss the 3D traced  by the metal
poor stars in the  bulge, with special emphasis on the spatial  distribution of RR Lyrae
(RRL) variables.
Section~\ref{sec:innerbar} presents  evidences for  the presence  of a  distinct
structure  within  a  radius  of $\sim$250  pc,  and  finally Section~\ref{sec:longbar}  reviews
evidences for  the presence of a  longer and thinner bar,  in addition to the  main one.

\section{THE MAIN GALACTIC BAR}
\label{sec:mainbar}

\subsection{Early evidences}  \label{early}

The   presence  of   a  bar   in   the  inner   Milky   Way  was   first  suggested   by
\cite{deVaucouleurs1964} as a  way to explain the departures from  circular motions seen
in  the  HI line  profile  at  21  cm, for  longitudes  close  to the  Galactic  center.
Nonetheless,  the first  direct evidence  of  the presence  of  a bar  was presented  by
\cite{blitz+1991}  who  analized  the  $2.4  \mu$m   maps  of  the  Galactic  center  by
\cite{matsumoto+1982}.   Curiously, these  authors  claimed that  the  Galactic bar  was
associated with the peanut shaped structure seen in the COBE maps by \cite{hauser+1990},
while  a  separate  and  more  metal  poor triaxial  spheroid  was  responsible  of  the
non-circular  motions in  the HI  line-of-sight  velocity versus  longitude maps.   Much
larger than the bar, their triaxial spheroid extended up to the solar radius.

After correcting the COBE  DIRBE maps by extinction it was soon  clear that the apparent
peanut  shape was  only  due  to a  dark  cloud,  known as  the  Pipe  Nebula, close  to
$(l,b)=(0,5)$  \citep{weiland+1994}.  The  boxy  structure, instead,  was confirmed  and
interpreted  as the  signature  of an  edge-on  bar, with  the near  side  in the  first
quadrant,  and the  major  axis at  an  angle $\theta=20^\circ  \pm  10^\circ$ with  the
Sun-Galactic  center direction  \citep{dwek+1995}.  Indeed,  the boxy  isophotes of  the
near-IR  COBE  maps   show  an  asymmetry  along  the  longitude   direction,  with  the
positive-longitude half being brighter than  the negative-longitude one.  Both the pivot
angle  and the  axis ratio  (1:0.3:0.2) measured  by \cite{dwek+1995}  are in  excellent
agreement with  the most recent  measurements based on  resolved stars (see  below). The
same authors assume that the whole bulge is bar-shaped, and derive the first photometric
mass of the  bulge/bar. From the total  $L_K$ luminosity and the  fuel consuption theory
they estimate the progenitor mass of the  post main-sequence stars, and then integrate a
Salpeter  IMF  along  the  main  sequence  down  to  0.1  M$_\odot$,  obtaining  M$_{\rm
  BULGE}=1.3\times10^{10}  {\rm  M}_\odot$.   The   total  $K$  luminosity  measured  by
\cite{dwek+1995}, $4.1\times10^8  L_\odot$, is  more than one  order of  magnitude lower
than the  value ($1.2\times10^{10}  L_\odot$) measured  by \cite{kent+1991},  using maps
from the  $SPACELAB$ infrared  telescope.  By accident,  however, the  bulge photometric
mass quoted  above is  remarkably similar  to the dynamical  mass derived  by \citet[][;
  $1.2\times10^{10}  {\rm   M}_\odot$]{kent+1992}  because  \cite{dwek+1995}   assume  a
Salpeter IMF  down to  0.1 M$_\odot$,  significantly steeper than  the observed  one, as
measured   with   near  IR   and   optical   star   counts  by   \cite{zoccali+00}   and
\cite{calamida+15}, respectively.

By  using  a different  and  3D  extinction correction  on  the  same COBE/DIRBE  data,
\cite{binney+1997}  also  constructed  a  photometric  model of  the  inner  Milky  Way,
confiming a pivot angle $\theta\approx 20^\circ$, for a bar with approximate axis ratios
(1:0.6:0.4), and  a pattern speed  of $\Omega_b=60-70$  km s$^{-1}$ kpc$^{-1}$.   A very
similar result ($\theta=25^\circ$ and $\Omega_b=50$ km s$^{-1}$ kpc$^{-1}$) was found by
\cite{fux1997, fux1999} by matching a self-consistent 3D N-body model with stars and gas
to the COBE DIRBE data (for stars) and the HI and CO $l-V$ diagrams (for the gas).

In the last  $\sim 10$ years, most of  the observational evidence for the  presence of a
bar in  the inner Milky  Way has been  gathered by studying  red clump (RC)  stars.  The
following section is dedicated to these kind  of studies. To be complete, however, there
are also other recent studies confirming the  existence of the main bar, using data from
microlensing  surveys  \citep[e.g.;][]{alcock+00}, or  OH/IR  and  SiO maser  kinematics
\citep{habing+06}.

\subsection{The bar from star counts}\label{RCcounts}

The most robust observational  evidence for the presence of a bar was  found by means of
RC stars used  as standard candles, in  order to de-project the  stellar distribution in
the inner  Galaxy. The first such  work was published by  \cite{stanek+94}, who analyzed
CMDs from  the Optical  Gravitational Lensing Experiment  \citep[OGLE;][]{udalski+92} in
three fields, one centered in Baade's  Window at ($l,b$)=($1,-3.9$), along the projected
minor axis, and the other two at ($l,b$)=($-5,-3.5$) and ($+5,-3.5$), respectively.  The
result was  that the  mean magnitude  of the red  clump (RC)  at positive  longitudes is
$\Delta V_{\rm RC}\sim 0.2$  mag brighter than that in Baade's Window,  which in turn is
$\Delta V_{\rm RC}\sim 0.2$ brighter than that at negative longitudes.  By assuming that
mean age and metallicity of the bulge stellar population does not have a smooth gradient
in longitude, this result was interpreted as  a smooth variation of the mean distance to
the bulge, with  longitude, e.g., the presence of a  bar.  \cite{stanek+94} measurements
were consistent with a bar pivot angle $\theta =45^\circ$.


These   early  results   were   confirmed   by  \cite{stanek+97,   lopez-corredoira+97,
  lopez-corredoira+00, nikolaev+97, unavane+98, bissantz+02, babusiaux+05, benjamin+05,
  nishiyama+05}. All these  studies used RC star counts to  constrain some triaxial bar
model.   The Besan\c{c}on  group \citep[e.g.][]{picaud+04,  robin+12} have  coupled star
  counts across  the whole  CMD with a  population synthesis models.   They also  fit a
  density model deriving  the bar pivot angle and scalelengths.   The latter parameters
  are somewhat  different in  all the  different studies quoted  above, with  the pivot
  angle in  particular ranging  from $10^\circ$ to  $30^\circ$, with  occasional values
  close  to $40^\circ$,  the  latter most  likely  influenced by  the  presence of  the
  so-called {\it long bar} (see \S\ref{sec:longbar}).

A special mention in this context deserves the work by \cite{rattenbury+07} who fitted a
triaxial bar density model to the observed RC distribution in 44 fields from the OGLE-II
survey.  With an area coverage about one order of magnitude larger than previous studies
($\sim11\,deg^2$), their best fitting  pivot angle is 24-27$^\circ$, the semimajor
and semiminor axis scalelengths in the plane ($x_0$ and $y_0$ respectively) and vertical
scalelengths  ($z_0$)  are  ($x_0:y_0:z_0$)=($1.2:0.4:0.3$) kpc  corresponding  to  axis
ratios ($10:3.5:2.6$).   This results  have been  partially confirmed  by \citet{cao+13}
using the final data release of OGLE-III \citep{soszynski+11}, which covers a much larger
area ($\sim90\,deg^2$). The  triaxial model that best fits the  RC distribution is found
with  adopting  a  slightly  larger  pivot angle  ($29^\circ  \--  32^\circ$)  and  axis
scalelength ($x_0:y_0:z_0$)=($1.00:0.41:0.38$) kpc.
 
\begin{figure}
\begin{center}
\includegraphics[width=\columnwidth]{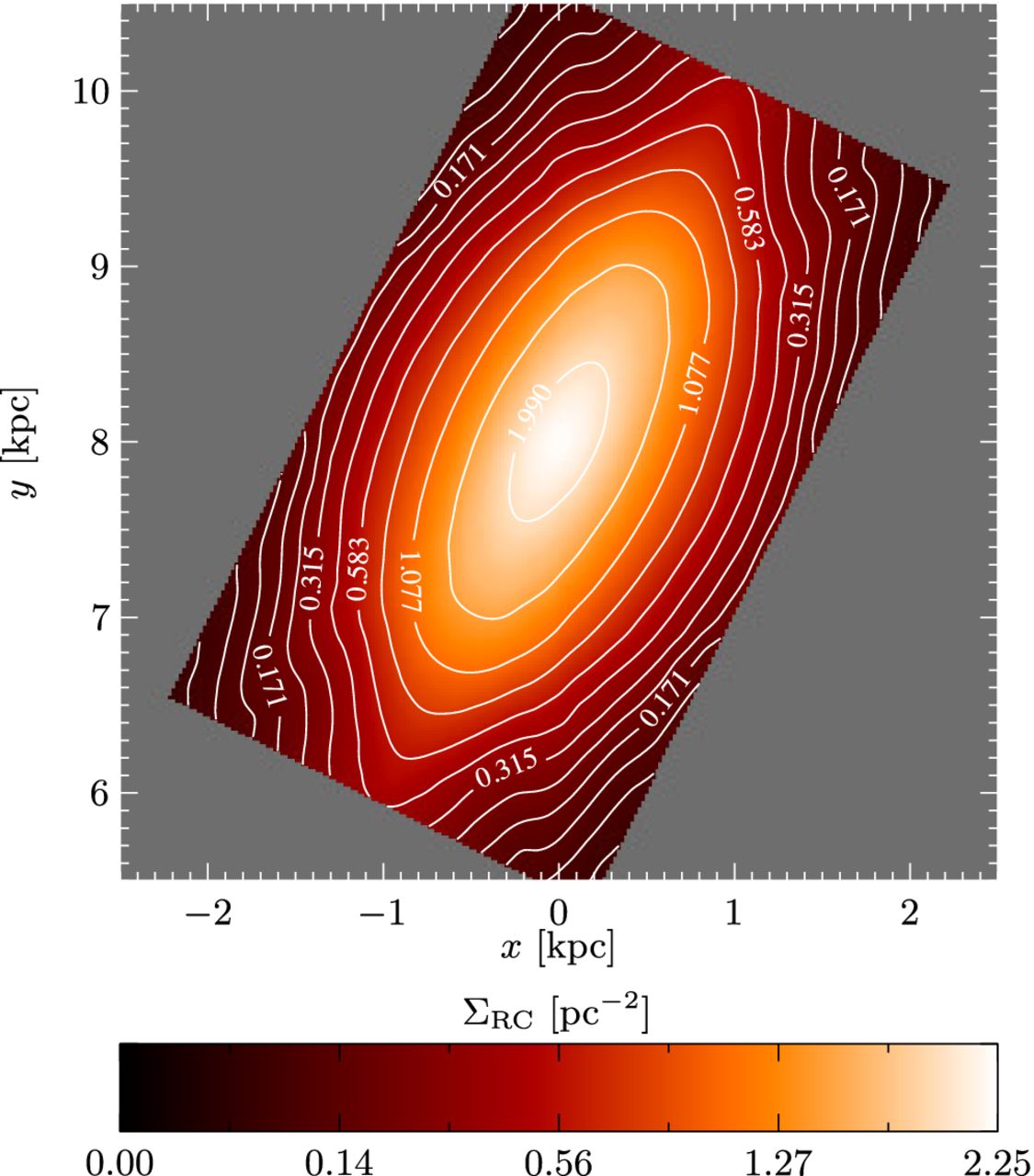}
\caption{The Galactic bar as seen from the North Galactic Pole. Numbers give the surface
density of RC stars in pc$^{-2}$, contours define isophotes separated by 1/3 mag. Figure
reproduced from \citet[][{\it Mapping the three-dimensional density of the Galactic
bulge with VVV red clump stars}, their Fig.17]{wegg+13}.
}\label{wegg1}
\end{center}
\end{figure}

More recently  the near-IR  VISTA Variables in  the V\'\i a  L\'actea ESO  Public Survey
\citep{minniti+10}  allowed  to map  the  whole  bulge  area within  $|l|<10^\circ$  and
$-10<b<+5$ with unprecedented accuracy.  The first data release \citep{saito+12} allowed
\cite{wegg+13}  to map  RC stars  across the  inner $2.2\times1.4\times1.1$  kpc of  the
Galactic bar.   Their best fitting model  has pivot angle of  $27^\circ\pm2^\circ$, axis
ratios   (10:6.3:2.6)   and   exponential   scalelengths   (0.70:0.44:0.18)   kpc.    In
Fig.~\ref{wegg1} we  reproduce Fig.~17  from \cite{wegg+13}, showing  the bar  as viewed
from above the plane  (note that the $x,y$ axes in this Figure  do not coincide with the
standard X,Y cartesian Galactic coordinates).

\section{THE X-SHAPE}
\label{sec:xshape}

The presence of a  double RC, i.e., with a split in magnitude,  in some fields along the
projected  minor axis  ($l=0^\circ$),  in  the outer  bulge  ($|b|>5^\circ$), was  first
noticed by  \cite{mw10-rio} and  \cite{zoc10-rio} in  several different  datasets.  This
result   was   confirmed   by   \cite{nataf+10}  using OGLE-II photometry,   and   by
\cite{mcwilliam+10} who interpreted  this feature, and the trend of  the RC magnitude in
the outer  bulge as evidence for  the outer bulge being  X-shaped. This was not  seen in
previous studies because they were all confined  to lower latitudes.  One year later the
RC magnitude  and density  was mapped  across the whole  bulge area  by \cite{saito+11},
using 2MASS photometry \citep{skrutskie+06}. They  confirmed the X-shaped structure seen
in the outer bulge, and also demonstrated that this feature was only the outer extension
of the main bar.  In other words, the two edges of the Galactic bar flare up in two {\it
  lobes}, forming a pronounced peanut, as seen in many external galaxies.

In fact,  the boxy  morphology is  characteristic of  all barred  galaxies seen  edge on
\citep{laurikainen+14}.  The peanut  shaped (or  X-shaped) structure  is also  a natural
product of bar  evolution, because dynamical instabilities produce  bending and buckling
of the elongated stellar orbits within the bars,  resulting in the shape of a peanut, or
an     X-shape    when     it    is     more    pronounced,     when    seen     edge-on
\citep{patsis+02,athanassoula05}.

\begin{figure}
\begin{center}
\includegraphics[width=\columnwidth]{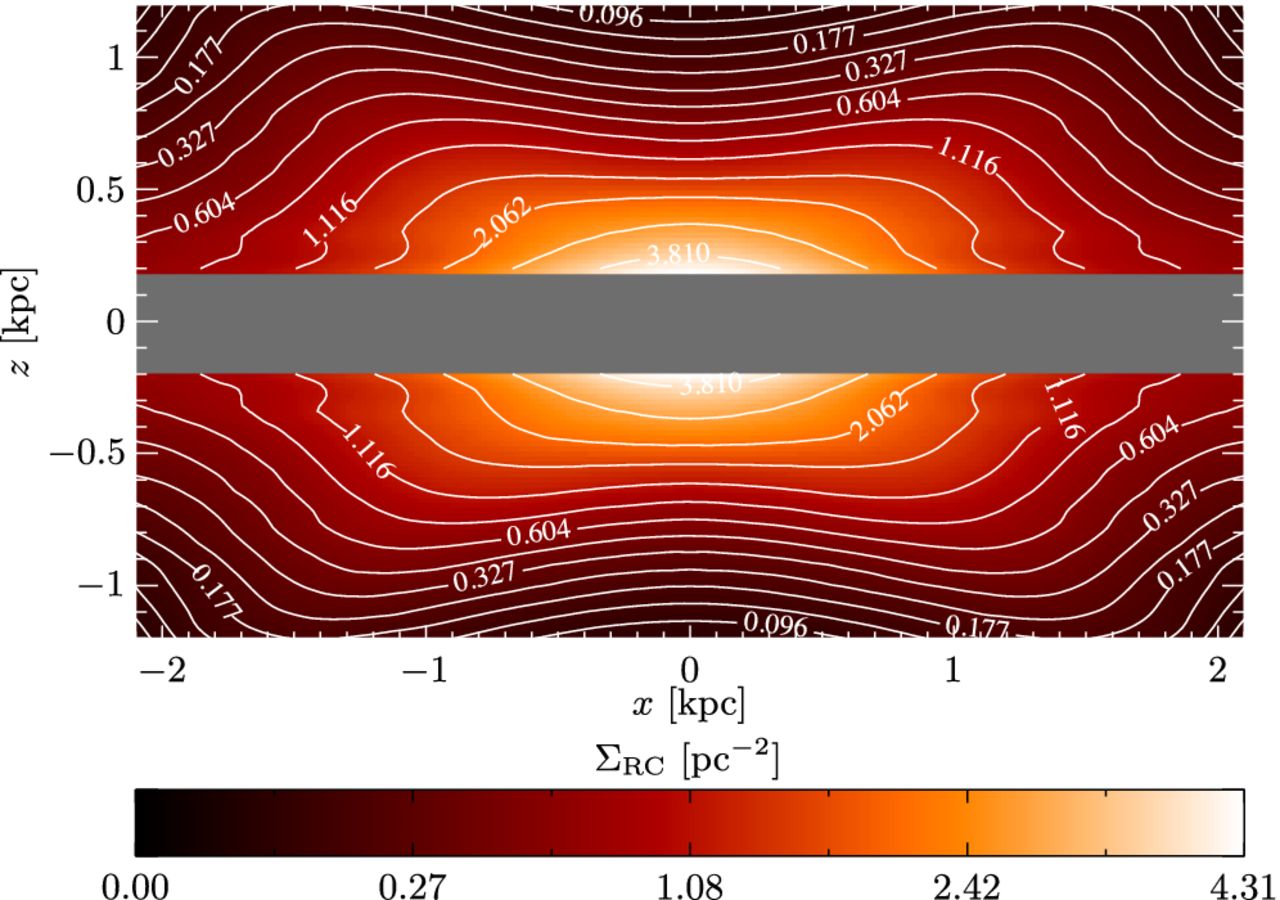}
\caption{The three-dimensional density of the Milky Way bulge measured in this work
projected along the intermediate axis. Numbers give the surface density of RC stars
in pc$^{-2}$, contours define isophotes separated by 1/3 mag. The extinction within
150 pc from the Galactic plane is too high for reliable density measurements, and is
therefore excluded from the projection.  Figure reproduced from \citet[][{\it Mapping 
the three-dimensional density of the Galactic bulge with VVV red clump stars}; their 
Fig.18]{wegg+13}.
}\label{wegg2}
\end{center}
\end{figure}

Due to a  brighter limit magnitude, the 2MASS photometry did not  allow to map the
regions closer to  the Galactic plane, and  RC star counts  were highly incomplete
already at $b\sim3^\circ$.  This also prevented a proper normalization between the inner
and outer density, affecting the assessment of the X-shape relevance with respect to the
main bar.  This problem  was overcome  thanks to the  VVV survey.   Indeed the  study by
\cite{wegg+13} mentioned above also included a  proper map of the outer, X-shaped bulge.
This is shown in Fig.~\ref{wegg2}, reproduced  from that paper, where their best fitting
bar model  is shown  edge on, and  the shape  of the outer  isophotes clearly  shows the
x-shape, or peanut shape.

An alternative  explanation of  the double  RC has been  proposed by  \cite{lee+15}, who
claimed that the density  and magnitude variation of the RC across  the bulge area could
also be explained  by the presence of a  spheroid with two populations of  stars, one of
which  is  helium enhanced,  similar  to  Terzan~5 \citep{ferraro+09,massari+14}.   This
spheroid would  therefore show  a RC  split in  magnitude.  However  this would  only be
visible  in the  outer bulge,  because in  the inner  bulge it  would be  erased by  the
presence of a bar, with a single RC population.  The main limitation of this scenario is
that it cannot  explain the absence of  the double RC at $|l|>2^\circ$,  i.e., away from
the  projected minor  axis, in  the outer  bulge, as  discussed in  a rebuttal  paper by
\cite{gonzalez+15}.
\section{THE METAL POOR SPHEROID}
\label{sec:spheroid}

By analyzing the  correlation between kinematics and metallicity for  K giants in three
fields along the  bulge projected minor axis, \cite{babusiaux+10}  first suggested that
the kinematics of the metal poor stars ([Fe/H]$\lesssim$0) was consistent with those of
a  spheroid, while  metal rich  stars ([Fe/H]$\gtrsim$0)  would have  elongated motions
typical of  galactic bars.  More specifically,  metal rich stars have  a non-negligible
vertex deviation  ($l_v\sim -40$) in  Baade's Window, while  the metal poor  stars have
$l_v\sim 0$.  In addition, the radial velocity dispersion of the metal poor stars would
stay roughly constant with distance from the  plane, while the dispersion of metal rich
stars  would be  significantly higher  closer to  the plane.   They thus  suggested the
presence of  two different populations  in the direction  of the Galactic  bulge.  This
suggestion was reinforced by \cite{hill+11} who show  how the metal rich and metal poor
components have different [Mg/Fe] distribution.

An  independent   confirmation  came  from  \cite{argosIII}   who  derived  metallicity
distribution function  of $\sim 14,000$ bulge  giants in 28 fields  across $|l|<30$ and
$5<|b|<15$, within the ARGOS survey \citep{argosII}. Thanks to the relatively large
magnitude range of their targets, across the RC, they could demonstrate that close to
the minor axis, only metal rich stars would show the split RC indicative of the X-shape.
Metal poor stars, on the contrary, would have a single RC. This demonstrates that 
metal poor bulge stars not only have different kinematics but also a different spatial
distribution.  This result was confirmed by \cite{rojas-arriagada+14}, using spectroscopic
data from the Gaia ESO Survey \citep{gilmore+12,randich+13}.

\cite{dekany+13}  used  RRL  from  OGLE-III  and  VVV  to  trace  the  oldest  (and
comparatively metal poor) component of the Galactic  bulge, and also found that they are
arranged in a spheroid with no trace of  a bar, nor obviously an X-shape. Same thing was
found  by \cite{catchpole+16}  using  Mira variables.   \cite{pietrukowicz+15}  on  the
contrary, using RRL from the OGLE-IV  survey \citep{soszynski+14} do confirm the presence
of a  bar in their spatial  distribution, although the extension,  ellipticity and pivot
angle of the structure  traced by RRL are all significantly smaller  than that traced by
RC stars.

\section{THE INNER BAR}
\label{sec:innerbar}

About    one    third   of    barred    galaxies    contain   secondary    inner    bars
\citep{laine+02,erwin11}, whose properties, such  as orientation, barycenter and pattern
speed  affect  the  gas  dynamics  of galaxies  \citep{rodriguez+08}.   Hence,  a  clear
knowledge  of the  morphology of  the  innermost bulge  regions, and  in particular  the
possible presence  of a secondary  smaller bar is relevant  to our understanding  of the
Galactic formation and evolution.

The question whether the Milky Way is a double\--barred galaxy is still debated. In what
follows we briefly  review some of the  studies that over the past  decade addressed the
issue about the existence of an inner bar  in the Milky Way. However, it should be noted
that this  possible {\it  inner bar} is  a structure whose  size ($\sim$1\,kpc)  is much
larger   than   the   so-called   {\it    nuclear   disk}   kinematically   defined   by
\citet{schonrich+15}.

The presence in the Galactic central region ($R\sim300$pc) of an inner bar nested inside
the main  bar was first esplicitly  suggested by \citet{alard01}. The  subtraction of an
exponential  profile, associated  with the  main  bar, from  the de\--projected  stellar
density map,  obtained through 2MASS star  counts, showed large residuals  in the region
$|l|\leq2^\circ$ and $|b|\leq2^\circ$. The longitudinal  elongation and asymmetry of the
derived residuals  in the  innermost region  could be  reproduced by  a small  bar, with
steeply dropping  density near its  edge. Therefore,  after ruling out  an inappropriate
extinction correction  and/or a substantial  deviation of  the main bar  density profile
from an  exponential distribution, as  possible cause  of the asymmetric  residuals, the
author  concluded that  in addition  to the  main  bar, an  inner and  smaller bar  with
different orientation might exist.

Later on,  \citet{rodriguez+08} used the Alard's  stellar density map to  constraint the
stellar mass  distribution adopted in their  gas flow dynamical model,  which included 3
components: disk, bulge and a nuclear bar  corotating with the main bar (i.e.  both bars
rotate  with  the same  speed).  Their  simulation reproduces  the  longitude\--velocity
diagram of  the Central  Molecular Zone as  the effect  of the nuclear  bar on  the gas,
excluding a possible lopsidedness  of the stellar potential due to  the nuclear bar. The
best\--fit  model is  found  for  a nuclear  bar  of mass  $(2\--5.5)\times10^9M_\odot$,
oriented by  an angle  of $\sim60^\circ\--75^\circ$ with  respect to  the Sun\--Galactic
Center line.

However, the  question whether the  density profile of the  central Milky Way  region is
axisymmetric and its  implication for the possible presence of  innermost structure date
back   to    earlier   time,   although   admittedly    controversial.   Indeed,   small
non\--axisymmetric  structures were  found  by  \citet{unavane+98} as  a  result of  the
comparison between  star counts in L\--band  observed in two different  fields along the
Galactic plane  at $l=\pm2.3^\circ$.  In  contrast with this  result, \citet{vanloon+03}
found that the dereddened luminosity function of  point sources at 7$\mu$m in the region
at $R\leq1$\,kpc are extremely symmetric around the bulge minor axis, thus suggesting an
azimuthally symmetric spatial distribution of the stellar population.

\begin{figure}
\centering
\includegraphics[width=7 cm]{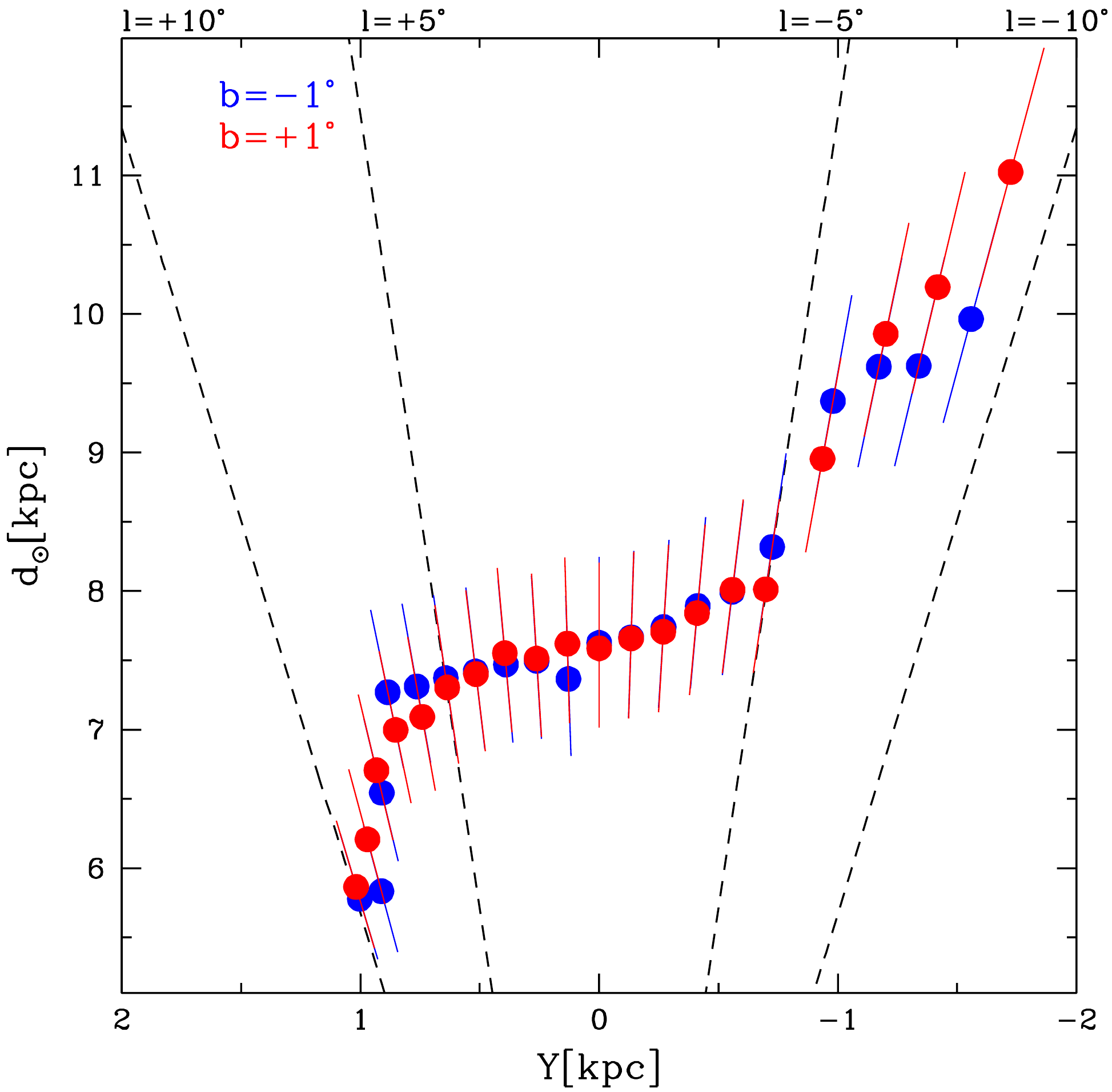}
\caption{Position of the  Galactic bar with respect  to the Sun traced by  the RC stars,
  and assuming an absolute magnitude $M_K=-1.55$ \citep{salaris02}. Blue and red circles
  respectively show the results for VVV data at latitudes $b=-1^\circ$ and $b=+1^\circ$,
  as derived by using the  PSF\--fitting photometric catalogs from\citet{valenti+15}. As
  in Fig.3 of  \citet{gonzalez+11a}, the solid lines identify the  distance spread along
  each  line of  sight correcting  for an  intrinsic bulge  dispersion of  0.17\,mag and
  photometric  errors.  Dashed  lines  refer  to   the  line  of  sight  for  longitudes
  $l=\pm5^\circ, \pm10^\circ$.
\label{fig1_inbar}}
\end{figure}

By using dereddened  color\--magnitude diagrams in the ($K_s$\ vs. $H-K_s$) plane to infer
the mean  magnitude of the RC  peak in the region  $b=-1^\circ$ and $|l|\leq10.5^\circ$,
\citet{nishiyama+05} observed  a clear change in  the slope of the  RC peak longitudinal
distribution.   The  observed overall  variation  of  the  RC  peak mean  magnitude  was
consistent with the  presence of a main  bar whose nearest edge is  oriented at positive
Galactic latitude (see  \S\ref{RCcounts}), however the shallower profile  in the central
region $|l|\leq4^\circ$ was interpreted as signature  of a distinct inner structure with
different  orientation angle.   The change  in  orientation of  the bar  in the  central
region, as traced by the RC population, was later confirmed by \citet{gonzalez+11a}.  By
using  VVV photometry\footnote{DR1  obtained through  aperture photometry},  the authors
extended the study of \citet{nishiyama+05} at latitude $b=+1^\circ$ finding an excellent
agreement.  Figure~3 reproduces the  result of \citet[][see their Fig.\,3]{gonzalez+11a}
by  using new  and more  accurate VVV  catalogs obtained  with PSF\--fitting  photometry
\citep[see][for further details]{valenti+15}. The figure shows  a clear change in the RC
profile distribution in the innermost region, $R\leq500$ pc.
\begin{figure*}
\centering
\includegraphics[width=15 cm]{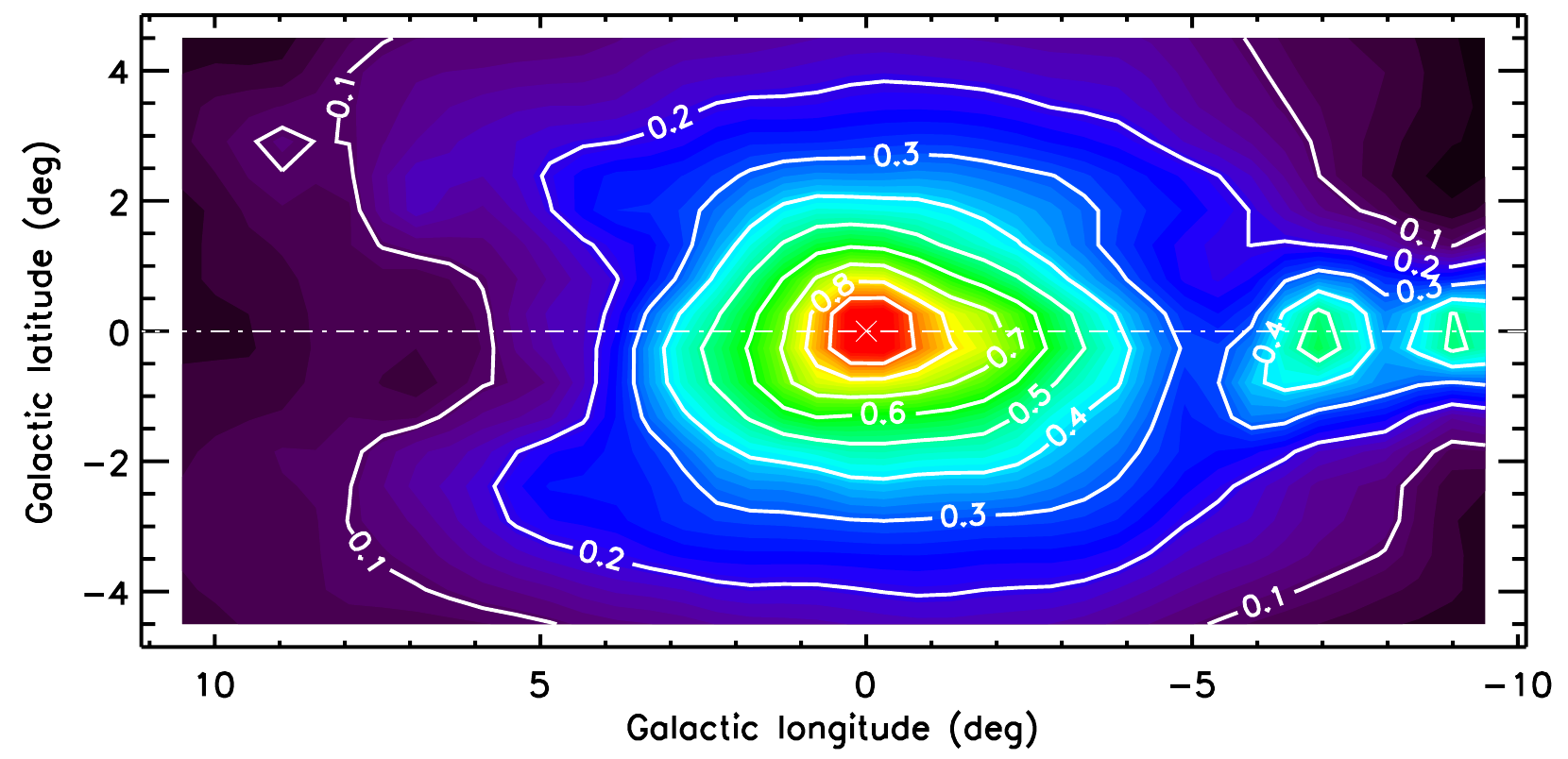}
\caption{Density map in  the longitude\--latitude plane based on RC star
  counts from \citet{valenti+15}. Star counts have  been normalised to the Maximum ({\it
    Max}).  Solid contours  are isodensity  curves, linearly  spaced by  0.1$\times${\it
    Max}\,deg$^{-2}$. 
\label{fig2_inbar}}
\end{figure*}

\begin{figure*}
\centering
\includegraphics[width=15 cm]{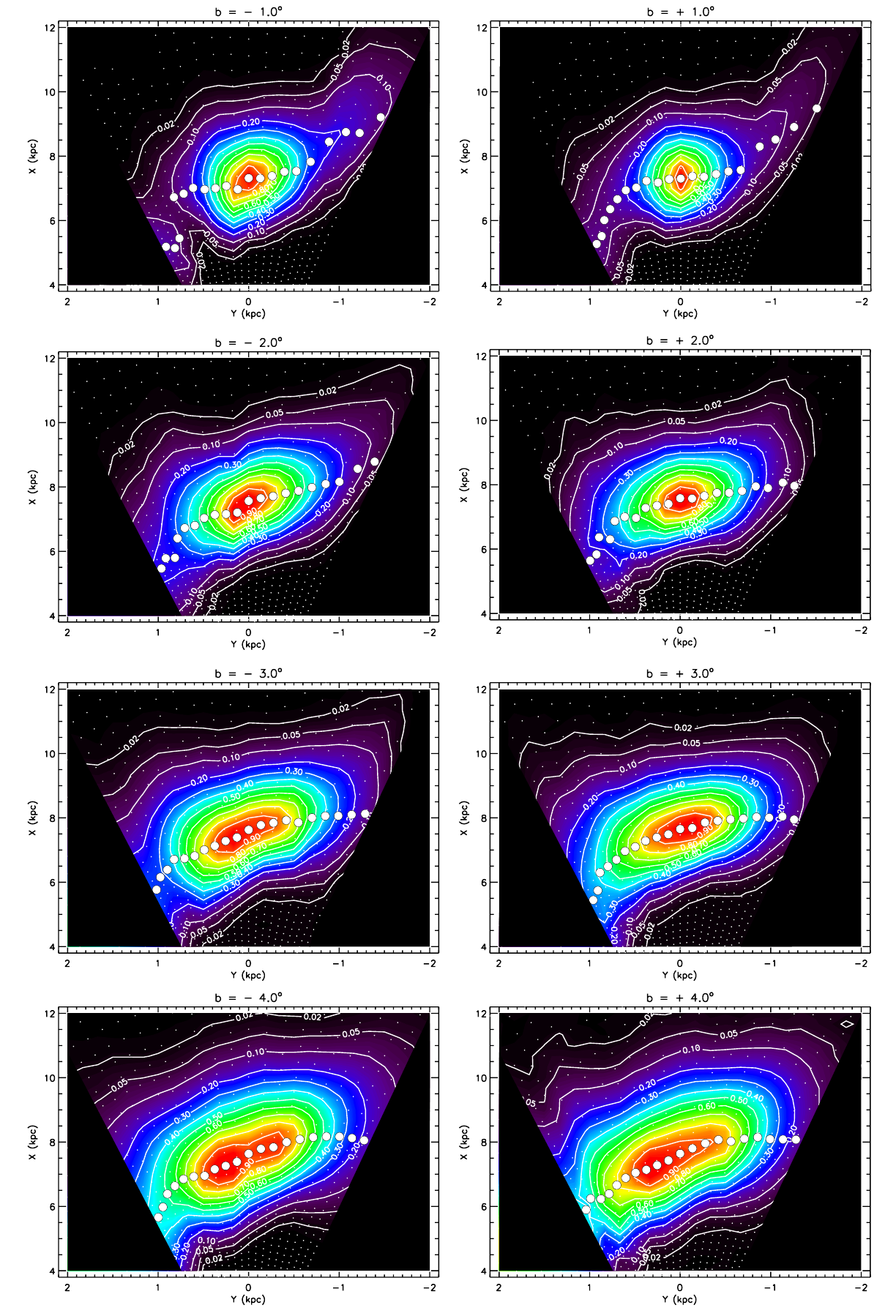}
\caption{De\--projection of the density map
  shown in Fig.\ref{fig2_inbar} at different latitude. Filled big circles identify the position
  of   the   RC  peak,   whereas   small   dotted   lines   refer  to   the   directions
  l=$1^\circ\,,\pm2^\circ\,,\pm3^\circ\,,\pm4^\circ\,,\pm5^\circ\,,\pm6^\circ\,,\pm7^\circ\,,\pm8^\circ\,,\pm9^\circ\,,\pm10^\circ$,
  as seen from an observer position.
\label{fig3_inbar}}
\end{figure*}


Being detected by  independent teams using different dataset  and extinction correction,
the change  of the  observed RC  profile slope along  the Galactic  plane in  the region
$|b|\leq1^\circ$ is  currently widely  accepted. Its interpretation  as signature  of an
inner bar  is, however, still  debated. In  this contest, the  N\--body model of  a boxy
bulge and bar  (with no inner bar) presented by  \citet{gerhard+12} nicely reproduce the
RC profile  observed by \citet{gonzalez+11a}.   The observed  variation in the  RC slope
would then be the result of a change  in the stellar density distribution along the line
of sight, i.e.  from highly elongated to nearly axisymmetric  isodensity contours in the
innermost regions.

Very recent  observational evidences seem  to support this  thesis. In fact,  as clearly
evident from the  VVV\--based RC stellar density profile  from \citet{valenti+15}, shown
here in Fig.\,\ref{fig2_inbar}, there is a  high\--density peak in the innermost region
(i.e.   $|l|\leq1^\circ, |b|\leq1^\circ$).   In  the observed  map  the central  density
contours  are  slightly  asymmetric  towards  negative  longitude.   However,  when  the
projection   effects  along   the  lines   of  sight   are  taken   into  account   (see
Fig.\,\ref{fig3_inbar})  the  isodensity  contours  become
progressively less  elongated when  moving closer  to the  Galactic center  (compare for
instance the shape of the isodensity contours at $b=\pm4^\circ$ and $b=\pm1^\circ$).  In
addition, a  tight correlation  between the  observed VVV density  map and  the velocity
dispersion  profile based  on the  GIBS  \citep{zoccali+14} kinematics  survey has  been
found,  such that  the $\sigma$\--peak  matches the  position of  the high  density peak
\citep[see Fig.3 of][]{valenti+15}.  This seems to confirm that the flattening of the RC
mean magnitude profile  in the region $|l|\leq4^\circ$  is more likely the  result of an
inner compact and axisymmetric spheroid rather than the presence of an inner bar.

\section{THE LONG BAR}
\label{sec:longbar}

Similar to  the boxy/peanut structures,  the formation of a  long bar (i.e.   planar bar
continuation)    is    also   a    common    outcome    of   bar    secular    evolution
\cite{athanassoula+05}. As is the case for  the main bar, a detailed characterisation of
the long  bar (i.e.   length, orientation,  pattern speed)  provides crucial  insight to
unveil the formation history of our  Galaxy. Indeed, numerical modelling have shown that
the bar  can affect the local  stellar velocity distribution of  the solar neighbourhood
\citep{dehnen00,minchev+10a}, it  can be  responsible for  stellar migration  and mixing
\citep{minchev+10b} in the disk, and for the observed non\--circular gas flow inside the
solar circle \citep{bissantz+03}.

The  presence of  strong peaks  along the  Galactic plane  at $l\sim33^\circ,  21^\circ,
27^\circ$  and  $-22^\circ$ in  the  near\--IR  large\--scale surface  brightness  maps,
available since  the late 70's,  stimulated a number  of studies aimed  at understanding
their origin  \citep[e.g.][]{hayakawa+81, okuda81, melnick+87}. This,  together with the
growing perception that the Milky Way was  a barred galaxy (see \S\,\ref{early}), led to
the very first study  in which the existence of a long bar  was suggested. Combining the
COBE/DIRBE surface brightness  map with RC star counts  from TMGS\footnote{ Two\--Micron
  Galactic   Survey   \citep{garzon+93}}  K\--band   photometry,   \citet{hammersley+94}
demonstrated  that  the  most likely  explanation  for  the  presence  of the  peaks  at
$15^\circ\le l \le35^\circ$ in the Galactic plane  is that they are associated with star
forming region at the near  end of a bar.  Their best\--fit model is  found for a bar of
semimajor  axis  R=3.7\--4\,kpc  and   orientation  angle  $\theta=75^\circ$.   Although
admittedly the \citet{hammersley+94} paper never explicitly mentioned the term {\it long
  bar}, the structure suggested  and modelled in that work will  later on be universally
termed as the {\it long bar}.

Later  on, \citet{hammersley+00}  traced the  long  bar by  looking at  the old  stellar
population selected from  near\--IR color\--magnitude diagram obtained  in regions along
the Galactic plane at $l=5^\circ,10^\circ,  15^\circ, 20^\circ, 27^\circ, 32^\circ$. The
differential  star  counts  were  found  to  be fairly  similar  across  the  fields  at
$27^\circ\leq l \leq 15^\circ$,  however the distance from the Sun of  the peak of stars
smoothly  increased with  decreasing longitude.   On the  other hand,  in the  innermost
observed  fields at  $l=5^\circ$ and  $10^\circ$, where  the bulge  contribution becomes
important, the star  counts increased considerably. The authors concluded  that the peak
at $l=27^\circ$ could not belong to the  bulge, but rather to a different structure such
as a  long bar, that  runs into the  bulge. Taking into  account the clustering  of very
young  stars  at  $l=27^\circ$  and  $21^\circ$  found  by  \citet{hammersley+94},  they
suggested that  the only  component that  can reasonably  explain all  the observational
evidences   is   a  bar   with   half\--length   $R\sim$4   kpc   and  a   pivot   angle
$\theta\sim43^\circ$.    These   results   have    been   substantially   confirmed   by
\citet{lopez-corredoira+01} based on  DENIS and TMGS star counts map  in two off\--plane
regions at $|b|\approx1.5^\circ$ and $|l|\ge30^\circ$.   In addition to the bar position
angle ($40^\circ$) and the semi\--major axis  ( 3.9\,kpc), they provided a more detailed
description  of its  properties.  The  closest edge  of the  bar is  found in  the first
quadrant at $l=27^\circ$ at a distance of  5.7\,kpc from the Sun (assuming a distance to
the Galactic center $R_0=7.9$\,kpc),  whereas the far end is in the  third quadrant at a
distance of 11.1\,kpc and is seen as the spur extending from the bulge to $l=-14^\circ$.
The  bar scale  height, as  traced by  the young  stellar population,  is about  50\,pc,
although the old component traced by \citet{hammersley+00} might have larger scale.

The further advent of  accurate and deeper surveys in the near  and mid\--IR such 2MASS,
GLIMPSE, UKIDSS and VVV has literally promoted a burst in the study of the Galactic long
bar, allowing  detailed investigations  on the  presence of  star counts  asymmetry over
homogeneously sampled larger scales.
Based on different surveys, independent teams found very similar results constraining to
very narrow  ranges the  bar angle  ($43^\circ\--45^\circ$) and  half\--length (3.7\,kpc
\---   4.4\,kpc)   \citep[see][]{benjamin+05,  lopez-corredoira+07,   cabrera-lavers+07,
  cabrera-lavers+08,  vallenari+08,  churchwell+09,   amores+13}.   Moreover,  the  vast
majority of these studies favour the thesis for  which the Milky Way hosts two bars: the
{\it  main bar}  confined  in  the bulge  within  $|l|\leq10^\circ$  with typical  angle
$\theta\sim25^\circ \-- 30^\circ$ (see \S\ref{RCcounts}),  and the {\it long bar} tilted
by $\sim45^\circ$ with respect to the Sun-Galactic Center line.  This configuration of 2
in\--plane misaligned bars has  been also detected in 3 (out of  6) external galaxies by
\citet{compere+14}, after  performing a  3D decomposition  modelling, on  K\--band 2MASS
images, including 3 different components: a disk, a bar and a triaxial bulge.

However, this picture has been very  recently challenged by \citet{wegg+15} who provided
a global  view of the Milky  Way bulge and  long bar by  using RC stars as  distance and
density tracers. Combining  UKIDSS, VVV, 2MASS and GLIMPSE data  they obtained a density
map of  the central $|l|<40^\circ$  and $|b|<9^\circ$, which is  best fitted by  a model
requiring an  orientation angle  of the long  bar consistent with  that of  the triaxial
bulge (i.e $28^\circ \-- 33^\circ$). The scale  height as traced by the RC stars changes
smoothly from the bulge  to the long bar.  They found evidence for  two scales height in
the long bar, suggesting the presence of a $\sim180$ pc thin bar component whose density
decreases outwards, and a $\sim45$\,pc  {\it superthin} component whose half\--length is
$\sim4.6$\,kpc. According to  the authors, the thin bar could  be the barred counterpart
of the solar neighbourhood  thin disc, whereas the superthin one  could be associated to
younger stars ($\sim1$\,Gyr) towards the end of the bar.

In agreement  with N\--body simulations  \citep{martinez-valpuesta+11, romero-gomez+11},
their findings support the scenario in which the  long and main bar are two parts of the
same structure, and that the Milky Way contains a central boxy/peanut bulge which is the
vertical extension of a longer, flatter bar.

On the other hand, the presence of  two in\--plane bars, with the long component twisted
with respect  to the barred  bulge is hardly predicted  by dynamical models  because two
separate rotating bars  should align with each other through  dynamical coupling in less
than a  few rotation  periods. In  this framework,  \citet{martinez-valpuesta+11} showed
that by  using a dynamical model  of a single stellar  bar evolved from the  disk, and a
boxy bulge originated  from it, through secular evolution and  buckling instability, can
reproduce the observations. In particular, the observed mismatch between the orientation
of the long ($45^\circ$) and main ($27^\circ$)  bar is caused by a combination of volume
effect  and  variation   of  the  density  distribution  along  the   observer  line  of
sight.  However,  regardless  from  the   problem  of  the  current  N\--body  dynamical
simulations to reproduce the possible two  bar configurations, the puzzling fact remains
that,  although  using the  same  tracers  (i.e.  RC  stars), methodology,  and  dataset
(i.e.  UKIDSS)  \citet{cabrera-lavers+08}  and \citet{wegg+15}  derived  very  different
results.

\section{SUMMARY AND DISCUSSION}

In this review we provided a description of the 3D structure of the Milky Way bulge from
observational perspectives.

Over the years, since the pioneering work of \citet{deVaucouleurs1964}, there has been a
large number of studies addressing the problem of the existence of the bar, and aimed at
characterising its main properties.  By using  a variety of techniques, ranging from the
integrated photometry  and star counts to  gas kinematics and microlensing,  these works
contributed to  build and strengthen  the general consensus that  our bulge is  indeed a
bar. The  near side of the  bar is in the  first Galactic quadrant, and  its orientation
with respect to the Sun\--Galactic Center line of sight is $\sim27^\circ$.

The advent  of IR surveys  (e.g. COBE/DIRBE, 2MASS and  VVV), allowing to  uniformly map
large scale area (e.g. $\geq 300\,deg^2$),  has finally provided a comprehensive view of
the Milky Way bulge  as a whole revealing its boxy/peanut  structure. This morphology is
typical of bulges formed  out of the natural evolution of  edge\--on barred galaxies, as
consequence of  disk instabilities and  bar vertical buckling. Moreover, the observed
magnitude split of the  RC in the outer region of  the bulge, is interpreted as evidence
of an X\--shaped structure, i.e., a pronounced boxy/peanut shape, which the models 
explain as the result of bar growing.

The  extensive  studies of  the  innermost  region of  the  bulge  (i.e. $|l|\leq2$  and
$|b|\leq2$) unveiled the presence of high axisymmetric stellar density peak. The latter,
rather than the presence of an inner bar,  seems to be responsible for the change in the
bar pivot  angle in this  region.  Moreover, the central  density peak matches  the peak
found in the stellar radial velocity dispersion.

Several studies focussing on  the region along the Galactic plane,  and outside the main
bar, disclosed  the existence of  a long bar, with  semimajor axis of  $\sim4.7$\,kpc in
length, and misaligned with  respect to the bulge main bar.  However,  the scenario of a
triaxial bulge with two in\--plane bars with different orientation as been challenged by
a very recent new results that, instead, suggests that the long bar is just the extension
at higher longitude of the bulge main bar.

The  correlation between  metallicity and  kinematics of  bulge giants,  as well  as the
distribution  of  RRL variables  suggest  the  presence in  the  Milky  Way bulge  of  a
metal\--poor  spheroid. However,  while the  general properties  in term  of morphology,
kinematics  and chemical  content of  the  boxy/peanut structure  have been  extensively
studied,  the  characterisation   of  the  metal\--poor  spheroids  is   far  for  being
complete. Additional  detailed investigation aimed  at confirming whether or  not giants
and RRL variables trace different  structure, and characterising the  kinematics of
the variables population, is still needed.

\begin{acknowledgements}
We  thank Chris  Wegg  for providing  Figs.\ref{wegg1} and  \ref{wegg2}, and  the
editors for inviting us  to write this review. MZ gratefully  acknowledge support by the
Ministry of  Economy, Development, and  Tourism's Millennium Science  Initiative through
grant IC120009, awarded  to The Millennium Institute of Astrophysics  (MAS), by Fondecyt
Regular 1150345, by  the BASAL-CATA Center for Astrophysics  and Associated Technologies
PFB-06 and by CONICYT's PCI program through grant DPI20140066.
\end{acknowledgements}

\bibliographystyle{apj}
\bibliography{pasabiblio}


\end{document}